# Nonequilibrium Statistical Operator for Systems of Finite Size

V.V.Ryazanov

Institute for Nuclear Research, 47, Nauky Prosp., Kyiv 03068, Ukraine

**Abstract**

The lifetime of statistical system is introduced. It is supposed that the nonequilibrium statistical operator implicitly contains the lifetime. The operations of taking of invariant part, averaging on initial conditions used in works of D.N. Zubarev, temporary coarse-graining (Kirkwood), choose of the direction of time are replaced by averaging on lifetime distribution. The expression for lifetime change at transitions from quasi-equilibrium system to nonequilibrium one is derived. A sort of the nonequilibrium statistical operator for systems of the finite sizes is suggested.

**Contents**

1.      **Introduction**

In the theory of random processes the lifetime of system is defined as time up to the moment of the first achievement of the certain (zero) level by random process *q(t)*, i.e.

$$\Gamma_x = \inf \{t: q(t)=0\}, \quad q(0)=x>0 \,. \tag{1}$$

Let's assume that this process *q(t)* describes the behaviour of some macroscopic physical value which is essential to the description of system (for example, order parameter). The lifetime is defined as random process subordinated (in sense of subordination of the theory of random processes [1]) to random process *q(t)* describing behaviour of system (parameter of the order, rough macroscopic variable,...). The characteristics of $\Gamma$ depend on *E=q(t)*. It is important to give physical interpretation of the mathematical definition (1). The lifetime of system is related to the time of steady existence of a system, that is to the time of its stay on a homeokinetic plateau whose distribution, for example, in work [2] is compared to entropic and information characteristics of system far from equilibrium, its response to external and internal influences, to stability and adaptive abilities of system. The states of a system on a homeokinetic plateau are characterized by mutual compensation of entropic effect related to the dissipation of energy, and by negentropic effect due to the presence of negative feedback. When a system exits the limits of the homeokinetic plateau there arise unstable structures and sharp qualitative changes in behaviour of system are observed. The value of lifetime $\Gamma$ is related to the exit time from a set of potential minima [3] and, accordingly, is influenced by temperature, external noises and is determined by the factors similar to those responsible for the values of relaxation and correlation times. The lifetime is also connected to the Kramers problem [4] of trespassing a potential barrier. Only for a certain class of systems "the lifetime" can be understood as their "vital time", i.e. the horizon of time through which no information of system as separate formation can pass. The macroparameters to which the lifetime "is subordinated" can be of various character. Generally "the lifetime" represents a consequence of existence of events in "history" of stochastic process playing so essential role in evolution of system, that they generate a flow of time for this system. These events change stochastic properties of process, being detected by an external observer, giving the information on system. It is known [5] what decisive role in the theory of random processes is played by a stopping times, an example of which the lifetime is. The coincidence of lifetime of system to time before the complete termination of their existence depends on force of internal interactions in system, its ability to form structure. The interaction with environment causes changes in the value *E=q* which are also reflected in the changes of $\Gamma$.

The existence and finiteness of lifetime value is ensured with presence of stationary states which physically correspond to existence of stationary structures. Complex functional and hierarchical interrelations between real systems correspond to the relations between lifetimes determining evolutionary processes as a sequence of transitions between various classes of states of system. Contrary to the traditional representation about time as something changeable, the lifetime acts as result of existence of stable stationary structures. It depends on external influences on system and internal interactions in it. Let's emphasize that lifetime is observable and measurable quantity.

Definition of lifetime and exit rate distinct from definitions (1) of the present paper are given in work [6], where it is done at the microlevel of phase space. But qualitative picture gained in works [6-7] corresponds to representations of the present work. Thermodynamic systems in which the transition from an initial state to final one occurs not infinitely slowly (as in equilibrium thermodynamics), but rather during some finite fixed time were considered in works [8-9].

The extremely important and interesting problem is the relation between time and information. In work [10] the identity of statistical thermodynamics and theories of the information is emphasized. Appearing as thermodynamic variable, lifetime (and the time in general) acquires information-theoretic sense. The detailed research of these phenomena should result in deeper understanding of degradation, destruction, dissipation, aging, death processes. In physics the lifetime usually is understood as time of life of the excited states (nuclei, power levels etc.), i.e. time of existence of random states which are distinct from equilibrium. In our approach lifetime is understood as complete lifetime, including equilibrium and nonequilibrium states.

In works of the Bruxelles school [11, 12] the constructive role of irreversibility is emphasized, the fact that nonequilibrium can be a source of the order. In [11] by consideration of a problem of Being and Becoming the operator of internal time is introduced. It is possible to find an analogy between it and lifetime. The initial conditions thus arise as outcome of previous evolution of system. "The Time is a key to understanding nature" [11].

The real systems have finite sizes and finite lifetime, that essentially influences their properties and properties of their environment, but is not reflected, for example in thermodynamics. The lifetime is represented by fundamental value having a dual nature. It is connected to the time flow, and with properties of a system as well. It is possible to classify systems by their ability to recycle, that is by the ability to be restored after the moment of degeneration (gases, liquid) or absence of such ability (solid states). The adequate mathematical model for the description of systems with finite lifetime can be served by stochastic models of a storage [13], which can be considered as models of systems, stimulated by the generalized noise. In them the interaction with thermostat, and moments of degeneration and stationarity loss are taken into account at certain conditions. This mathematical technique is generalized on a case of time-dependent factors and the characteristic times are compared with relaxation and correlation times. Analogously the lifetimes can be applied to the description of large-scale fluctuations resorption, comparing with irreversible processes, and to a number of other phenomena.

Both lifetime and internal time from [11] serve for the description of properties of dissipative systems with inherent age, and choosing of a direction of time flow. The lifetime is determined by trajectories with the allocated point of the moment of achievement of a zero level in the future. The approach by I. Prigogine [11] and the concept of lifetime have in common the presence of some allocated moment, event, as the inalienable parts of the probabalistic description and birth of new (among already present systems, elements, interactions). In the present work it is the moment of destruction of system, and in [11] - bifurcation, transition to chaos, as a consequence of instability, i.e. other pole of stationary – relaxation evolutions of system described, for example, by the model of a storage [13, 14]. Let's note that the description suggested here supposes natural generalization on cases of loss of stability, transition to states among which there are no stationary ones [14]. The times before transition in these states are also determined which are precisely given by a storage model.

The dual nature of lifetime having property of both dynamic variable and external coordinate was already mentioned above. Its latter feature corresponds to Newton's classical conception of absolute empty space as a vessel for a matter. In the concept of lifetime these representations are united with ideas about the matter which generates time, meanwhile the time without substance (or field) does not exist. The existence of time is resulted by the existence of complex systems formed from elements, driven by influence forces. There is no absolute time, and there are times of specific systems.

In thermostatic it is supposed that any isolated thermodynamic system with time flow comes in a equilibrium state and never spontaneously leaves from there. However assumption of an opportunity of fluctuations in system (i.e. random deviations of internal parameters from their equilibrium values) contradicts the basic principles of thermostatics [15]. The assumption of stay of system in basically (equilibrium) state for infinitely long time will be replaced by more physical assumption (appropriate also to statistical description) about an opportunity of the exit of system from this state and destruction, degeneration of system, under influence of internal fluctuations.

## 2.   Nonequilibrium Statistical Operator Method and Lifetime of System

Nonequilibrium statistical operator method (NESOM) [16-25] turned to be an effective tool in the solving the nonequilibrium problems. On its basis the information statistical thermodynamics is formulated [26-33]. In works [26-33] the assumptions made when introducing NESOM are in details described.

In work [21] NESOM is interpreted as choosing of the solution set of the Liouville equation appropriate to the reduced description, through averaging on the initial moments of time. The solution to the Liouville equation, satisfying to the initial condition

$$\rho(t)_{/t=t_0} = P\rho(t_0) = \rho_q(t_0,0) , \qquad (2)$$

where $P$ is projection operator [21], $\rho_q(t_0,0)$ is quasi-equilibrium distribution (9) [20-21], has a form

$$\rho(t,t_0) = \exp\{-iL(t-t_0)\}\rho_q(t_0,0) ; \qquad (3)$$

$L$ is Liouville operator; in classical case $iLA = -\{H,A\} = \Sigma_k((\partial H/\partial p_k)(\partial A/\partial q_k) - (\partial H/\partial q_k)(\partial A/\partial p_k))$; $H$ is the system Hamiltonian, $p_k$ and $q_k$ are pulses and coordinates of particles; $\{...\}$ is classical Poisson's bracket. The solution (3) contains non-physical dependence on the initial moment of time $t_0$ and has physical sense at $t \gg t_0$, when initial conditions become insignificant. The assumption is done that the evolution of system with equal probability can begin from any initial condition in an interval from $t_0$ up to $t$, which is considered large enough. The averaging (3) on the initial moments of time gives expression

$$\rho(t) = \int_{t_0}^{t} \exp\{-iL(t-t_0)\}\rho_q(t_0,0)dt_0/(t-t_0) . \qquad (4)$$

The time $t-t_0$ should be big enough for dampfing of initial nonphysical states and formation of necessary correlations. From (4) is one gets

$$\partial\rho/\partial t + iL\rho(t) = -(\rho(t)-\rho_q(t,0))/(t-t_0) , \qquad (5)$$

i.e. Liouville equation contains a source, which tends to zero at $t_0-t \to \infty$. This source breaks symmetry to the reversibility of time, defines boundary conditions to the Liouville equation and selects required, i.e. retarded solutions. Using equality of limits following from the common theory of Tauber theorems, in [21] integral in (4) is replaced with expression

$$\rho(t) = \int_{-\infty}^{0} (1/T)\exp\{t_1/T\}\exp\{iLt_1\}\rho_q(t+t_1,0)dt_1 =$$

$$\int_{-\infty}^{t} w(t,t_0)\rho_q(t_0,t_0-t)dt_0 , \qquad (6)$$

where $w(t,t_0)=(1/T)\exp\{(t_0-t)/T\}$; $T=\varepsilon^{-1}$; $\exp\{-iL(t-t_0)\}\rho_q(t_0,0)=\rho_q(t_0,t_0-t)$. In [20, 21, 29] the function $w$ are treated as a kernel of integrated transformations. Having done in (6) the transformation of variable $t=t_0+y$, $t_1=-y$, and having written down (6) for $ln\rho$, we shall receive

$$ln\rho(t)=\int_0^\infty p_q(y)ln\rho_q(t-y,-y)dy \ . \tag{7}$$

where $w(t,t_0) \to p_q(y)=\varepsilon exp\{-\varepsilon y\}$, i.e. instead of auxiliary weight function from [29] it is suggested to use density of distribution of value $t-t_0=y$, which is interpreted as lifetime of system (random value), described by quasi-equilibrium distribution. Thus, the initial moment $t_0$ is replaced on $t-y$ with averaging on $y$. The value of lifetime can be limited both from above and from below. Then the integration at averaging will be carried out not from $0$ up to $\infty$, as in (7), but rather from the minimal value of lifetime $\Gamma_{min}$ up to maximal $\Gamma_{max}$. For $\rho(t)$ of the kind (7) the expression of a source in the Liouville equation changes which for $p_q(y)=\varepsilon exp\{-\varepsilon y\}$; $\varepsilon=1/T$, and takes on the form

$$\partial\rho/\partial t+iL\rho(t)=-\varepsilon\ (\rho(t)-\rho_q(t,0)) \ . \tag{5`}$$

A source in this equation, as well as in the equation (5) has the form of relaxation member describing relaxation $\rho(t)$ to quasi-equilibrium distribution with average time $T=\varepsilon^{-1}$.

Mathematical operations similar to (7), or

$$ln\rho(t)=\int_{0^+}^\infty p_q(y)ln\rho_q(t-y,-y)dy=ln\rho_q(t,0)-\int_{0^+}^\infty (\int p_q(y)dy)[dln\rho_q(t-y,-y)/dy]dy \tag{8}$$

(where the partial integration is carried out at $\int p_q(y)dy_{|\ y=0+}=-1$; $\int p_q(y)dy_{|\ y\to\infty}=0$; at $p_q(y)=\varepsilon exp\{-\varepsilon y\}$; $\varepsilon=1/T$, the expression (8) passes in the nonequilibrium statistical operator (NESO) [20, 21]), can be found in [34] and in [1], where it is shown, how it is possible to construct from random process $\{X(t)\}$, appropriate to evolution of quasi-equilibrium system, the set of new processes, introducing the randomized operational time. It is supposed that to each value $t>0$ there corresponds a random value $\Gamma(t)$ with distribution $p^t_q(y)$. The new distribution of random value $X(\Gamma(t))$ is defined by equality (8). The random values $X(\Gamma(t))$ form new random process, which, generally speaking, need not to be of Markovian type any more. Each moment of time $t$ of "frozen" quasi-equilibrium system is considered as random value $\Gamma(t)$ of the finishing of lifetime with distribution $p^t_q(y)$. Any moment of lifetime can be the last moment of life with a certain probability. Since the interval $t-t_0=y$ was large enough (see paragraph before the formula (4)), it is possible to introduce the minimal lifetime $\Gamma_{min}=\Gamma_1$ and to integrate in (7)-(8) on the interval $(\Gamma_1,\infty)$. It results in the change of the normalization density of distribution $p_q(y)$. For example, the function $p_q(y)=\varepsilon exp\{-\varepsilon y\}$ will be replaced by $\varepsilon exp\{\varepsilon\Gamma_1-\varepsilon y\}$, $y\geq\Gamma_1$; $p_q(y)=0$, $y<\Gamma_1$. The lower limit of integration in (7)-(8) by $\Gamma_1\to 0$ is equal to $0^+=lim_{c\to 0}(0+c)$. The lower limit of integration in Laplace transformation ((15) type) is equal to $0^-=lim_{c\to 0}(0-c)$. Thus: $p_q(y=0^-)=\varepsilon exp\{-\varepsilon y\}_{/y=0^-}=0$; $p_q(y=0^+)=\varepsilon$.

One of conditions of applicability of algorithm of phase integration of complex systems [35], which result is limiting exponential distribution $p_q(y)=\varepsilon exp\{-\varepsilon y\}$, is the long enough stay of system in a class of states. If one ignores details in evolution of system, i.e. does not take into account transitions of system inside a separate class of states, provided that the system long enough stays in each class, the dependence of transitions between classes from evolution of system inside classes disappears, and the time of stay in each class becomes the sum of large (random) number of random values - times of stay in separate states, that at the natural initial assumptions allows to consider as distributed by the exponential law. The exponential distribution for lifetime is also obtained in work [36], where large times are also assumed. It is possible to choose $p_q(y)=Cf(y)$, $y<t_1$; $p_q(y)=\varepsilon exp\{-\varepsilon y\}$, $y\geq t_1$; $C=[1-exp\{-\varepsilon t_1\}](\int_0^{t_1} f(y)dy)^{-1}$. It is possible to take the function $f(y)$ from models of the theory of queues or from other sources estimating lifetime distribution for small times [1, 34]. According to results of work [36] we choose the value of $t_1$ from the condition $t_1>>\sigma^2(x)/K$. The value $\varepsilon$ can be defined from results of work [36]. If to simulate the behaviour of quasi-equilibrium system by the Fokker-Plank equation $\partial\omega_q(x)/\partial t=\partial[\omega_q\partial U/\partial x]/\partial x+K(\partial^2\omega_q/\partial x^2)/2$ and by the random process $x(t)$, $\dot x =f(x)+\xi$; $<\xi>=0$; $<\xi\xi_\tau>=K\delta(\tau)$; $f(x)=-dU(x)/dx$, where $U$ is potential, $N=\int exp\{-2U(x)/K\}dx$, $x_{st}$ is stationary values of parameters $P_j$ of quasi-equilibrium system (9), the correlation function $<xx_\tau>=\sigma^2 R(\tau)$, $1/\varepsilon=2(N/K)\int_{x_{st}}^{0} exp\{2U(x)/K\}dx$. Used in works [20-21] exponential density of probability of lifetime $p_q=\varepsilon exp\{-\varepsilon y\}$ is limiting distribution for lifetime in the theory of phase integration of complex systems [35] for a case of one class ergodic states. In work [16] the uniform density of probability is used, which is not good attested mathematicaly. Many results of the probability and random processes theory can be used for a definition $p_q(y)$. So, there can be also amendments taking into account subsequent terms of asymptotic expansion series, except for the basic limiting expression in exponential probability [35]. In a case $n$ of classes ergodic states in system the limiting exponential distribution is replaced by generalized Erlang distribution (for $n=2$: $p_q(y)=\theta\varepsilon_1 exp\{-\varepsilon_1 y\}+(1-\theta)\varepsilon_2 exp\{-\varepsilon_2 y\}$). The expressions for density of probability of lifetime dependent on a history of system can be used [37], as well as the relations of the renewal theory and density of probability of lifetime of non-failure operation etc. Detailed consideration of functions $p_q(y)$ requires a separate investigation (e.g., [56]). We note only that generally it is possible to replace $\varepsilon exp\{-\varepsilon y\}$ by $p_q(y)$, and $\int \varepsilon exp\{-\varepsilon y\}dy=-exp\{-\varepsilon y\}$ - by $\int p_q(y)dy$. In the present work we shall restrict ourselves by exponential distributions.

It is essential that $\varepsilon \neq 0$. The thermodynamic limiting transition is not performed, and actually important for many physical phenomena size dependence of system is considered. We assume $\varepsilon$ and $\Gamma_0$ to be finite values. Thus the Liouville equation for $\rho(t)$ contains a finite source, which is explained by the openness of system, its interaction with an environment and finiteness of system lifetime. The similar source is introduced in work [38]. Assumption about finiteness of lifetime breaks temporary symmetry. And such approach (introduction $p_q(y)$, averaging on it) can be considered as completing the description of works [20-21].

Primary in such approach there is an assumption of existence of random value of lifetime $\Gamma_q$ of quasi-equilibrium system of finite size, in which the microscopic sources of dissipative processes in system are not taken into account, but fluctuations are possible. Dynamic variable are random values and have finite lifetimes (at certain conditions [14, 39]), determined as (1). In the quasi-equilibrium distribution

$$ln\rho_q(z;t_1,t_2)=-\Phi(t_1)-\sum_{j=1}^{n}\int dx F_j(x,t_1)P_j(z/x,t_2) \qquad (9)$$

thermodynamic parameters $F_j$ obviously depend on time, and $P_j$ depend on time in a classical case through evolution operator, $P_j(z/\ t`-t)=exp\{i(t`-t)L\}P_j(z)$. In (9) $\{P_j\}$, $j=1,2...$, are a basic set of dynamical variables, dependent from $z$ - points in microscopic phase space, $\{F_j(x,t)\}$ are the set of Lagrange multipliers that the variational method introduces, and which are determined in terms of the basic macrovariables (average values) $Q_j(r,t)=Sp\{P_j(r)\rho_q(t,0)\}$, $x$ is spatial coordinate (here momenta can also be included), thermodynamic potential $\Phi$ is defined from a normalization condition;

$$\Phi(t)=lnSp\{exp[-\sum_{j=1}^{n}\int dx F_j(x,t)P_j(z/x)]\};$$ in [20] for local-equilibrium distribution [20-21] (for the description of a hydrodynamical stage of a nonequilibrium process): $P_0(z;x)=H(z;x)$; $P_1(z;x)=p(z;x)$; $P_{i+1}(z;x)=n_i(z;x)$; $F_0(x,t)=\beta(x,t)$; $F_1(x,t)=-\beta(x,t)v(x,t)$; $F_{i+1}(x,t)=-\beta(x,t)(\mu_i(x,t)-m_iv^2(x,t)/2)$. Here $H(z;x)$ is dynamic variable density of energy, $n_i(z;x)$ are density of particles for the $i$-th component, $p(z;x)$ is pulse density; $\beta(x)$ plays a role of local reverse temperature, $\mu(x)$ is the local chemical potential, $v(x)$ is the mass speed, $m_i$ are masses of $i$-th particles. If one defines NESO from a entropy maximum [20, 40] at restrictions, valid for all past moments of time $t_0 \leq t' \leq\ `t$, in [40] for Lagrange multipliers of NESO the expression takes on the form $\varphi_j(r;t`,t)=w(t,t`)F_j(r,t)$, that in interpretation of lifetime corresponds to proportionality of Lagrange multipliers of NESO to the lifetime probability density, i.e. physical condition of their belonging to finite information-gathering interval $(t_0,t)$ [29, 40], treated as lifetime of quasi-equilibrium statistical system.

In works [29, 32, 40] the analogy between integration on time in (6) and time-smoothing procedure [41] is marked. All evens in history of system are taken into account (like the lifetime concept does). The quasi-equilibrium statistical operator (9) describes macrostates of the system in a time interval, around $t$, much smaller than the relaxation times of the basic variables (implying in a "frozen" equilibrium or quasi-equilibrium in such interval). For larger time intervals the effects of the dissipational processes comes into action. The distribution function and the average value is composed of two contributions: one is the average with the quasi-equilibrium distribution (meaning the contribution of the state at the time *t*), plus the contribution arising out of the dynamical behavior of the system (the one that accounts for the past history and future dissipational evolution).

There are some interpretations of operation (6). So, in [20, 21, 29] it is treated as necessary for breaking the temporary symmetry, as in a Bogoliubov's method of quasi-averages [42], where the time symmetry breaking is reached by introduction of the appropriate source in the Liouville equation, which tends to zero after thermodynamic limiting transition. The lifetime of infinite system is also infinite. It also explains necessity of tending to zero of value $\varepsilon=1/\Gamma_0$ when the volume of system tends to infinity. For finite systems the values $\Gamma_0$ are finite also. In [20] the analogy to the formal scattering theory is considered. In [40] it is marked that the time-smoothing procedure introduces a kind of Prigogine's dynamical condition for dissipativity [11, 43]. The Liouville equation (5`) with a source has Boltzmann-Bogoliubov-Prigogine symmetry. Zubarev [21] considers a source in the Liouville equation as "thermostat" influence, due to which the phase point representing system, making free evolution, is thrown from one phase trajectory to another with probability *w(t,t₀)*. The reception of the information is related to the last history of the system macrostates along its evolution from an initial condition. One more treatment of expression (6) is related with the *fading memory* process which may be considered as the statistical-mechanical equivalent of the one proposed in phenomenological continuum-mechanical-based Rational Thermodynamics [44, 45]. In Zubarev's approach this fading process occurs in an adiabatic-like form towards the remote past: as time evolves memory decays exponentially with lifetime $\varepsilon^{-1}$. The evolution of the nonequilibrium system is influenced more strongly by recent correlations, that corresponds to the Bogoliubov's principle of correlations weakening. Thus irreversible behavior in the system is stated introducing in a peculiar way a kind of Eddington's time-arrow. In [46] an interesting alternative approach to the derivation of Zubarev's form of NESOM was presented. The authors of [46] suggest the turn on of an adiabatic perturbation for *t`>t₀*, instead of a fading-memory interpretation. A basis of those approaches is the fact that no real system can be wholly isolated. The Zubarev's procedure also appears as having certain analogies with so-called repeated randomness assumptions [47, 48] as discussed by del Rio and Garsia-Colin [49]. In Appendix IV [40] other alternative derivations are presented, which were used for obtaining the ensemble algorithm in equilibrium, also following the ideas proposed by McLennan [50], and a connection with an earlier proposal by I.Prigogine is discussed.

## 3. The Connection between of Quasi-equilibrium and Nonequilibrium Lifetimes in Markov Type Approach

In the present section it is shown that density of probability of lifetime of nonequilibrium system differs from density of probability of lifetime of quasi-equilibrium system.

The distribution of probabilities $\rho(z;t)$ in phase space $z=\{p,q\}$ of coordinates and pulses can be related to the density of distribution of macrovalues $P_n$ (see, for example, [38, 51, 52]). In [52] it is considered the Markov system (system is called of Markov type, if a Markov model can be constructed for it). Thus there should be such temporary scales, as constant macroscopic relaxation time $t_r$, indicating speed of change of the chosen thermodynamic parameters $P_n(z)$, and constant time $t_0$ of an establishment of local equilibrium or time of correlation of random influences for parameters $P_n(z)$. Thus $t_r >> t_c >> t_0$; $t_c=(t_r t_0)^{1/2}$. The density of distribution of small number of the chosen thermodynamic parameters $P_n(z)$, $n=1,2..., r$, is connected with $\rho(z)$ by relation

$$\omega(P) = \widehat{\Pi} \rho = \int \delta(P-P(z))\rho(z;t)dz , \qquad (10)$$

where $\widehat{\Pi}$ is linear operator of mapping of space of distributions in phase space on space of distributions $\omega$. In [52] the reverse operator $\widehat{\Pi}^-$ was also constructed. Let's note that more strict approach demands introduction of the generalized reverse operator determined in [35]. Two-temporary density of distribution is equal [52] to

$$\omega[P(t_2),P(t_1)] = \int \delta[P_2(t_2)-P(z)]exp\{L(t_2-t_1)\}\rho(z;t_1)\delta[P(z)-P_1(t_1)]dz ,$$

and density of probability of transition of Markov process is equal to conditional density of distribution

$$\omega[P(t_2)/P(t_1)] =$$
$$\int \delta[P_2(t_2)-P(z)]exp\{L(t_2-t_1)\}\rho(z;t_1)\delta[P(z)-P_1(t_1)]dz / \int \delta[P_1(t_1)-P(z)]\rho(z;t_1)dz . \quad (11)$$

If we insert (11) for quasi-equilibrium distribution (9), (11) takes a form

$$\omega_q[P(t_2)/P(t_1)] =$$
$$\int \delta[P_2(t_2)-P(z)]\rho_q(z;t_1,t_1-t_2)\delta[P(z)-P_1(t_1)]dz / \int \delta[P_1(t_1)-P(z)]\rho_q(z;t_1,0)dz . \quad (12)$$

The stochastic potential dependent on argument $v$ and random value $y(t_2)$ is defined in [52] by a relation

$$\Phi_{t_2}(v,y(t_2)) = \lim_{\tau \to 0} \tau^{-1} \int (exp\{\sum_{j=1}^{n} v_j(y_{j_{t_2+\tau}} - y_{j_{t_2}})\} - 1) \omega(\vec{y}(t_2+\tau)/\vec{y}(t_2)) d\vec{y}_{t_2+\tau} \quad (13)$$

also determines master-equation for the distribution density $p(y,t)=\omega(P(t)=y(t),t)$,

$$\partial p(y,t)/\partial t = N_{\partial,y} \Phi(-\partial/\partial y, y) p(y,t) , \quad (14)$$

(the operator $N_{\partial,y}$ defines the order of operations (differentiating on $y$ goes the last)) and also the equation for Laplace $L$ transformation from density of probability $p_0$ of time of the first achievement of border (zero) by process $y(t)$, which is taken as definition of lifetime (1),

$$L(k,y_0) = \int_{0^-}^{\infty} exp\{-kt\} p_0(t,y_0) dt; \; y_0 = y(t=t_0); \; L(k=0,y_0) = L(k,y_0=0) = 1 ; \quad (15)$$

$$kL(k,x) = N_{x,\partial} \Phi(\partial/\partial x, x) L(k,x); \; x = y_0 . \quad (16)$$

Substituting (12) (at $t_1-t_2=-\tau$, $t_1=t_0$) in (13), we shall get the stochastic potential $\Phi_q$ for quasi-equilibrium distribution. We shall consider how the value $\Phi$ will change at replacement of transitive density of probability (12) by the appropriate probability for nonequilibrium distribution (7) (or (8)), i.e. at replacement of $\rho_q$ in (12) by $\rho(t)$. First of all we shall note that for the logarithm of nonequilibrium distribution $ln\rho(t)$, given by equality (8), the equation (5`) is fair (after replacement $\partial/\partial t$ on $-\partial/\partial y$ and partial integration the rhs of (5`) is equal to $dln\rho(t)/dt$). Under the initial condition $\rho(t_0)=\rho_q(t_0,0)$ [21], if in (8) we assume that $\rho(t_0-y,-y)=0$ at $y>0$, as at the moment of time, smaller than $t_0$, the system does not exist. Therefore, if we replace in (12) $\rho_q(t_0,0)$ by $\rho(t_0)=\int_0^\infty \varepsilon exp\{-\varepsilon y\}\rho_q(t_0,0)dy=\rho_q(t_0,0)$, $\rho_q(t_0,-\tau)$ on $\int_0^\infty \varepsilon exp\{-\varepsilon y\}\rho_q(t_0,-\tau)dy=\rho_q(t_0,-\tau)$, the value (12) will not change. The moment $t_0$ is chosen, as in expressions (15) - (16) $x=y(t_0)=P(t_0)$, and for these relation in (13) one should set $t_2=t_0$. Let's carry out other replacement of probabilities. From the value of $\rho_q(t_0,-\tau)$ with fixed value $t_0$ we shall proceed to random value of the initial moment of time $t_0=t-y$ and make averaging on $y$ (or on $t_0$), i.e. (by consideration only of case $p_q(y)=\varepsilon exp\{-\varepsilon y\}$) we shall replace $\rho_q(t_0,-\tau)$ on the value

$$exp\{\int_0^\infty \varepsilon exp\{-\varepsilon y\}ln\rho_q(t-y,-\tau)dy\}=exp\{ln\rho_q(t,-\tau)+\int_0^\infty exp\{-\varepsilon y\}[dln\rho_q(t-y,-\tau)/dy]dy\}=$$

$$\rho_q(t,-\tau)+\rho`(t,-\tau); \quad \rho`(t,-\tau)=\sum_{k=1}^\infty [\int_0^\infty exp\{-\varepsilon y\}\dot{S}(t-y,-\tau)dy]^k \rho_q(t,-\tau)/k! ; \qquad (17)$$

$$\dot{S}(t-y,y)=dln\rho_q(z;t-y,y)/dy=-\sum_{j=1}^n (\partial F_j(t-y)/\partial y)[P_j-<P_j>^{t-y}]-\sum_{j=1}^n F_j(t-y)\dot{P}_j \qquad (18)$$

is the operator of entropy production [20, 21, 29-33]. It corresponds to absence of fixing of the initial moment of time and averaging on $t_0$ [21] (relations (4), (6); in the second argument $y=\tau$). Value $\rho_q(t_0,0)$ in a denominator of (12) we shall replace on $\rho(t,0)=\rho_q(t,0)+\rho`(t,0)$. Then $\omega_q[F(t_0)P(t_0+\tau)/F(t_0)P(t_0)]$ will be replaced on $\omega_q[F(t)P(t_0+\tau)/F(t)P(t_0)]$, and

$$\omega[P(t_0+\tau)/P(t_0)] = \{\omega_q[P(t_0+\tau)/P(t_0)] + A\}/(1-B), \qquad (19)$$

where

$$A = \int \delta[P_2(t_0+\tau)-P(z)]\rho`(z;t,-\tau)\delta[P(z)-P_1(t_0)]dz / \int \delta[P_1(t_0)-P(z)]\rho_q(z;t,0)dz; \qquad (20)$$

$$1-B = \int \delta[P_1(t_0)-P(z)]\exp\{\int_0^\infty \varepsilon\exp\{-\varepsilon y\}\ln\rho_q(z;t-y,0)dy\}dz / \int \delta[P_1(t_0)-P(z)]\rho_q(z;t,0)dz =$$

$$\rho[t;P(z)=P_1(t_0)]\omega(P_1)/\rho_q[t,0;P(z)=P_1(t_0)]\omega(P_1) = \exp\{\int_0^\infty \exp\{-\varepsilon y\}\sum_{j=1}^n \int dx[(\partial F_j(t-y)/\partial y)(<P_j>^{t-y}-P_j(t_0))-F_j(t-y)\dot{P}_j]dy\}; \quad \omega(P_1)=\int\delta[P_1(t_0)-P(z)]dz. \qquad (21)$$

We shall substitute (19) in (13), (16) at $t_2=t_0$, $y(t_2)=P(t_0)$, $v=\partial/\partial P(t_0)$:

$$\Phi = \lim_{\tau\to 0}\tau^{-1}\int[\exp\{(\partial/\partial P(t_0))(P(t_0+\tau)-P(t_0))\}-1][\omega_q/(1-B)]dP(t_0+\tau) +$$
$$\lim_{\tau\to 0}\tau^{-1}\int[\exp\{(\partial/\partial P(t_0))(P(t_0+\tau)-P(t_0))\}-1][A/(1-B)]dP(t_0+\tau). \qquad (22)$$

In the first term of the right part (22) let us act by the operator identity $f(\partial/\partial y)\exp\{vy\}=\exp\{vy\}f(\partial/\partial y+v)$ [52] to *(1-B)* as (21) and we shall note that *(1-B)* does not depend from $P(t_0+\tau)$ and it is possible to take it out from the integral on $P(t_0+\tau)$. In the second term (22) lets apply to (19) - (21) the operator identity $\exp\{(\partial/\partial P(t))(P(t+\tau)-P(t))\}\varphi(P(t))=\varphi(P(t+\tau))$ and following from (9) relation $\partial\rho_q[F(t)P(t)]/\partial P(t)=-\int d^3r \sum_{j=1}^n F_j(t)\rho_q[F(t)P(t)]$, relation $\int\rho`(z;t,-\tau)dz=0$, together with partial integration, having replaced $t_0$ on $t-y$ with averaging on $y$. The integration on $dP(t+\tau)$ "removes" $\delta$-functions, the complete operator $\lim_{\tau\to 0}(\tau)^{-1}\int[\exp\{(\partial/\partial P(t))(P(t+\tau)-P(t))\}-1](...)dP(t+\tau)$ at $\tau\to 0$ works as $[\partial\varphi(P(t))/\partial P(t)][\partial P(t)/\partial t]$. In a result we shall get:

$$\Phi(\partial/\partial x) = \Phi_q(\partial/\partial x)/(1-B) - D[1-1/(1-B)], \qquad (23)$$

where $D = \sum_{j=1}^{n} \int dx \int_{0}^{\infty} exp\{-\varepsilon y\}[(\partial F_j(t-y)/\partial y)<\dot{P}_j>^{t-y} + F_j(t-y)(\partial <\dot{P}_j>^{t-y}/\partial y)] dy =$

$\int_{0}^{\infty} exp\{-\varepsilon y\}[d<\sigma(r,t-y)>/dy] dy = \int_{0}^{\infty} p_q(y)[d\bar{S}(t-y)/d(t-y)] dy - d\bar{S}(t)/dt;\ Sp\{-\dot{\bar{S}}(r;t,0)\rho(t)\} =$

$<\sigma(r,t)> = \sum_{j=1}^{n} \int dx F_j(x,t) d<P_j(x,t)>/dt = d\bar{S}(t)/dt$ [20, 21, 29];

$<P_j(x,t)> = \int dz \rho_q(z;t,0) P_j(z/x) dx = \int dz \rho(z;t) P_j(z/x) dx;\ \bar{S}(t) = \int d^3r s(r,t) = -<ln\rho_q(t,0)>$ is the nonequilibrium (or *informational* [33]) entropy [20, 21, 29]. More general, but also more cumbersome form of record of value $<\sigma>$ is given in [33].

At $n=1$, $P=E=H$, $F=\beta$; $<\dot{P}_j>^{t-y} = <\dot{E}>^{t-y}$; $D = \int dx \int_{0}^{\infty} e^{-\varepsilon y}[-k_B \sigma_y \sigma_{t-y} T(y)(\partial T(t-y)/\partial T(t))/T(t-y) c_v(y) + (\partial <\dot{E}>^{t-y}/\partial y)/k_B T(t-y)] dy$; $\sigma_y = \beta(y)<\dot{E}>^y$; $c_v(y) = \partial <E>^y/\partial T(y)$; $\beta(y) = 1/k_B T(y)$, $k_B$ is the Boltzmann constant. It is possible to write down (23) and through correlators on quasi-equilibrium distribution, through which in [20] the relating average values of fluxes with thermodynamic forces are expressed by kinetic factors, and also through correlation function of the operators entropy production, through which in work [20] the entropy production is expressed. The values $D$ and $B$ in (23) are expressed through entropy fluxes and entropy production.

Derivative on time from thermodynamic variable $F_m$ (9) is expressed in [20] in approximation of hydrodynamics of an ideal liquid. More general expressions for $\partial F_m(t)/\partial t$ are given in [21]: $\partial F_m(t)/\partial t = \sum_{k=1}^{n} (\delta F_m(t)/\delta <P_k>^t)<\dot{P}_k>^t$; $(\delta F_m(t)/\delta <P_k>^t) = -(P,P)^{-1}_{km}$. Even more general expressions for this value are given in [40]:

$d\vec{F}(t)/dt = i\hat{\Omega}(t)\vec{F}(t) + \int_{0}^{\infty} dy exp\{-\varepsilon y\}\hat{\Gamma}(-y|t)\vec{F}(t-y)$, where $\hat{\Omega}(t) = i\hat{\bar{C}}^{-1}(t)(\vec{P},\dot{\vec{P}}|t)$, $\hat{\Gamma}(-y|t) = \hat{\bar{C}}^{-1}(t)(\dot{\vec{P}}[\hat{1}-P_\varepsilon(t)]exp\{-iyL\}[\hat{1}-P_\varepsilon(t)]\vec{P}|t)$, $\dot{\vec{P}} = (1/i\hbar)\{\vec{P},\hat{H}\}$, $P_\varepsilon(t)\hat{A} = \bar{C}^{-1}(t)(P,\hat{A}|t)$ is projective operator; $\bar{C}^{-1}$ is matrix, reverse to a correlation matrix with elements

$C_{jk}(t)=Sp\{\int_0^1 du\,\hat{P}_j[\rho_q(t,0)]^u \Delta\hat{P}_k[\rho_q(t,0)]^{-u+1}\};\quad \Delta\hat{P}_k=\hat{P}_k-Sp\{\hat{P}_k\rho_q(t,0)\};\quad \hat{1}$ is unit operator; $(\vec{P},\dot{\vec{P}}|t)=Sp\{\int_0^1 du\,\vec{P}[\rho_q(t,0)]^u \Delta\dot{\vec{P}}[\rho_q(t,0)]^{-u+1}\};\quad \Delta\dot{\vec{P}}=\dot{\vec{P}}-Sp\{\dot{\vec{P}}\rho_q(t,0)\}.$

As $\vec{I}_s(r,t)=\sum_{j=1}^{n}\vec{I}_j(r,t)F_j(r,t)$ is the entropy flux, in (23) there appears an entropy flux; $\partial\langle P_j\rangle^t/\partial t+div\,\vec{I}_j(r,t)=\sigma_j(r,t);\quad \partial s(r,t)/\partial t+div\,\vec{I}_s(r,t)=\sigma_s(r,t);\quad \sigma_s(r,t)=\sum_{j=1}^{n}\vec{I}_j(r,t)\nabla F_j(r,t)+\sum_{j=1}^{n}F_j(r,t)\sigma_j(r,t);\quad \vec{I}_j(r,t)$ are fluxes of variables $\langle P_j\rangle$, $s(r,t)$ is the local entropy density. It is possible to write down (23) and through entropy production.

From (16-17) and (23) is received, that

$$p(y|x)=C[p_q(y|x)]^{1/(1-B)}\exp\{-Dy[1-1/(1-B)]\},\qquad (24)$$

where $C$ is appropriate normalization. For $p_q(y|x)=\varepsilon\exp\{-\varepsilon y\}$,
$p(y|x)=\{[\varepsilon/(1-B)]+D[1-1/(1-B)]\}\exp\{-y\{[\varepsilon/(1-B)]+D[1-1/(1-B)]\}\};\quad \varepsilon=1/\Gamma_0;$

$$\langle\Gamma\rangle=\Gamma_0(1-B)/\{1-\Gamma_0 D[1-(1-B)]\}.\qquad (25)$$

The expressions for average on $\rho(t)$ of meaning $ln(1-B)$ are obtained in [29-33]. Averaging (21) and (25) on $\omega[P(t\to t_0=t-y)]$ (10), assuming, that $\langle\Gamma\rangle$ and $\Gamma_0$ do not depend from $P_1(t_0)$ [35]. Then (25) corresponds as

$$\langle\Gamma\rangle=\Gamma_0\exp\{-\Delta_\varepsilon\bar{S}(t)\}/[1-\Gamma_0 D(1-\exp\{-\Delta_\varepsilon\bar{S}(t)\})];\qquad (26)$$

$\Delta_\varepsilon\bar{S}(t)=\int dx\int_0^\infty e^{-\varepsilon y}Sp\{-\dot{S}(x;t-y,0)\rho(t_0=t-y)\}dy=\sum_{j=1}^{n}\int dx\int_0^\infty e^{-\varepsilon y}F_j(x,t-y)[d\langle P(x)\rangle^{t-y}/dy]dy=$

$\int_0^\infty e^{-\varepsilon y}[d\langle S(t-y)\rangle/dy]dy=\int_0^\infty p_q(y)\langle S(t-y)\rangle dy-\langle S(t)\rangle=-[D+d\bar{S}(t)/dt]/\varepsilon;\qquad (27)$

$\int_0^\infty p_q(y)\langle S(t-y)\rangle dy=\langle\bar{S}(t)\rangle_0;$

$$<\Gamma> = \Gamma_0 exp\{-\Delta_\varepsilon \overline{S}(t)\}/[1+[\Gamma_0 d\overline{S}(t)/dt+\Delta_\varepsilon \overline{S}(t)](1-exp\{-\Delta_\varepsilon \overline{S}(t)\})] =$$

$$\Gamma_0 exp\{\Gamma_0[D+d\overline{S}(t)/dt]\}/[1-\Gamma_0 D(1-exp\{\Gamma_0[D+d\overline{S}(t)/dt]\})] \ .$$

The value of $<\Gamma>$ decreases or increases in comparison with $\Gamma_0$, depending on a correlation between amount entropy produced in system and entropy flows between system and environment.

The similar result can be obtained from application (10) to (17). Differentiating this result on $t$ taking into account $\partial\omega/\partial t = N_{\partial_y}\Phi(-\partial/\partial y,y)\omega$, (14), we shall get: $\Phi(-\partial/\partial y,y) = \Phi_q(-\partial/\partial y,y)(\omega_q/\omega) + \omega^{-1}(P(t))\partial(\hat{\Pi}\rho')/\partial t$. At replacement of arguments in $\Phi(-\partial/\partial y,y)$ on $\partial/\partial x, x$ we shall replace and $\partial/\partial t$ on $\partial/\partial t_0 = (\partial/\partial P(t_0))(\partial P(t_0)/\partial t_0)$. Averaging the received result on (10), we obtain the result coinciding with (23).

### 4. Nonequilibrium Statistical Operator for Systems of Finite Sizes

In the present section it is considered, how it is possible to construct the nonequilibrium statistical operator for systems of finite volume. Nonequilibrium statistical system evolves to equilibrium (or to stationary state) as a rule. The chaotic states (for example, dynamic chaos) are generally possible. It is necessary to take into account fractal character of phase trajectories and effects due to an openness of system. In introduction it was marked that in equilibrium exist the fluctuations, and the system of the finite sizes also exists finite time. It can leave equilibrium (or stationary state) under the action of various sort of external influences. The probability of degeneration of system is equal to $P_0 = 1/Q$, where $Q$ is statistical sum. For systems of the finite sizes we assume that $<\Gamma> \neq \infty$, $\varepsilon > 0$.

We shall designate $\rho_q = \rho^{(0)}$; $\rho = \rho^{(1)}$; $\Gamma = \Gamma^{(1)}$. If the nonequilibrium function of distribution (in a classical case) $ln\rho^{(1)}$ is obtained by averaging of function $ln\rho^{(0)}$ on density of probability $p^{(0)} = \varepsilon exp\{-\varepsilon y\}$; $\varepsilon = k^{(0)} = 1/\Gamma_0$; $\Gamma_0 = <\Gamma^{(0)}>$, of quasi-equilibrium system lifetime $\Gamma^{(0)} = \Gamma_q$ (7), density of probability of lifetime $\Gamma^{(1)}$ of nonequilibrium systems with distribution $\rho^{(1)}$ can be written down as $p^{(1)}(y) = k^{(1)} exp\{-k^{(1)} y\}$, where $k^{(1)} = exp\{\Delta_\varepsilon \overline{S}^{(0)}(t)\}(k^{(0)} - D^{(0)}) + D^{(0)} = 1/<\Gamma^{(1)}>$, the values $D^{(0)} = D$, $\Delta_\varepsilon \overline{S}^{(0)}(t) = \Delta_\varepsilon \overline{S}(t)$ are given in (23), (27). And similarly to averaging of $ln\rho^{(0)}$ on $p^{(0)}$ (7) it is possible to carry out the operation of averaging of $ln\rho^{(1)}$ on $p^{(1)}$. Thus there is one more temporary smoothing in the sense of Kirkwood [41], where a history of system and details of its behaviour at the previous moments of time is specified. I.e., the relation similar (7), defines system with nonequilibrium function of distribution $\rho^{(2)}$, where

$$ln\rho^{(2)}(t) = \int_0^\infty p^{(1)}(y) ln\rho^{(1)}(t-y) dy = ln\rho^{(0)}(t,0) + \int_0^\infty exp\{-k^{(0)}y\}[dln\rho^{(0)}(t-y,-y)/dy]dy$$

$$+ \int_0^\infty exp\{-k^{(1)}y\}[dln\rho^{(0)}(t-y,0)/dy]dy + \int_0^\infty exp\{-k^{(1)}y\} \int_0^\infty exp\{-k^{(0)}y_0\}[d^2ln\rho^{(0)}(t-y-y_0,-y_0)/dydy_0]dy_0 dy = \int_0^\infty p^{(1)}(y) \int_0^\infty p^{(0)}(y_0) ln\rho^{(0)}(t-y-y_0,-y_0) dy_0 dy .  \qquad (28)$$

For system with this function of distribution average lifetime is also defined:

$$1/<\Gamma^{(2)}> = k^{(2)} = exp\{\Delta_\varepsilon \bar{S}^{(1)}(t)\}[k^{(1)} - D^{(1)}] + D^{(1)}, \qquad (29)$$

where $\Gamma^{(2)}$ is lifetime of system with function of distribution $\rho^{(2)}$,

$D^{(1)} = <<d\bar{S}(t)/dt>_0>_1 - <d\bar{S}(t)/dt>_0;$

$<<d\bar{S}(t)/dt>_0>_1 = \int_0^\infty p^{(1)}(y) \int_0^\infty p^{(0)}(y_0)[d\bar{S}(t-y-y_0)/d(t-y-y_0)] dy_0 dy;$

$\Delta_\varepsilon \bar{S}^{(1)}(t) = <<\bar{S}(t)>_0>_1 - <\bar{S}(t)>_0 = \int_0^\infty exp\{-k^{(1)}y\}[d<S(t-y)>/dy + \int_0^\infty exp\{-\varepsilon y_0\}[d^2<S(t-y-y_0)>/dydy_0]dy_0]dy;$ $<<\bar{S}(t)>_0>_1 = \int_0^\infty p^{(1)}(y) \int_0^\infty p^{(0)}(y_0) \bar{S}(t-y-y_0) dy_0 dy .$

It is possible to continue similar iterative procedure, having assumed

$$D^{(2)} = \int_0^\infty exp\{-k^{(2)}y\}[d(d\bar{S}(t-y)/d(t-y))/dy]dy + \int_0^\infty exp\{-k^{(2)}y\}\{d(\int_0^\infty exp\{-k^{(0)}y_0\}d[d\bar{S}(t-y-y_0)/d(t-y-y_0)]/dy_0)/dy\}dy_0 dy + \int_0^\infty exp\{-k^{(2)}y\}\{d(\int_0^\infty exp\{-k^{(1)}y_0\}d[d\bar{S}(t-y-y_0)/d(t-y-$$

$y_0)]/dy_0)/dy\}dy_0dy + \int_0^\infty exp\{-k^{(2)}y\}\{d(\int_0^\infty exp\{-k^{(1)}y_1\}d(\int_0^\infty exp\{-k^{(0)}y_0\}d[d\bar{S}(t-y-y_1-y_0)/d(t-y-y_1-y_0)]/dy_0)/dy_1)/dy\}dy_0dy_1dy = <<<d\bar{S}(t)/dt>_0>_1>_2 - <<d\bar{S}(t)/dt>_0>_1; \ldots$

$ln\rho^{(r)}(t) = \int_0^\infty p^{(r-1)}(y) ln\rho^{(r-1)}(t-y)dy; \quad p^{(r)}(y) = k^{(r)}exp\{-k^{(r)}y\}, \quad 1/<\Gamma^{(r)}> = k^{(r)} = exp\{\Delta_\varepsilon \bar{S}^{(r-1)}(t)\}$

$[k^{(r-1)} - D^{(r-1)}] + D^{(r-1)}; \quad D^{(r)} = <<\ldots<d\bar{S}(t)/dt>_0\ldots>_{r-1}>_r - <<\ldots<d\bar{S}(t)/dt>_0\ldots>_{r-2}>_{r-1};$

$\Delta_\varepsilon \bar{S}^{(r)}(t) = <<\ldots<\bar{S}(t)>_0\ldots>_{r-1}>_r - <<\ldots<\bar{S}(t)>_0\ldots>_{r-2}>_{r-1};$

$<\ldots<(\ldots)>_0\ldots>_r = \int_0^\infty p^{(r)}(y)\ldots\int_0^\infty p^{(0)}(y_0)(\ldots)dy_0\,dy.$

On any $f$-st stage the procedure stops at $k^{(f-1)} = k^{(f)}$, that occurs at $k^{(f-1)} = k^{(f)} = 1/<\Gamma^{(f-1)}> = 1/<\Gamma^{(f)}> = D^{(f-1)}$. The function $\rho^{(f)}$ also will be NESO for system of the finite size. Integrating $D^{(r)}$ and $\Delta_\varepsilon \bar{S}^{(r)}(t)$ by parts we shall find

$D^{(r)} = \int_{0^+}^\infty exp\{-k^{(r)}y\}[d<\ldots<d\bar{S}(t-y)/d(t-y)>_0\ldots>_{r-1}/dy]dy;$

$\Delta_\varepsilon \bar{S}^{(r)}(t) = \int_{0^+}^\infty exp\{-k^{(r)}y\}[d<\ldots<\bar{S}(t-y)>_0\ldots>_{r-1}/dy]dy.$

The use of the obtained distributions results in specification of the description. For example, in the expression of average values of fluxes through function of distribution $\rho^{(2)}$ there occur additional terms in comparison with similar expressions through function of distribution $\rho^{(1)}$ [20].

Lets write down $\rho^{(1)}$ as

$$\rho^{(1)} = Q^{-1} exp\{-\sum_m \int [F_m(x,t)P_m(x) - \int_0^\infty exp\{-\varepsilon y\}j^m(x,-y)X_m(x,t-y)dy]dx\}, \qquad (30)$$

which is obtained in [20] with the help of the equations of hydrodynamics of an ideal liquid and with application of thermodynamic relations in the local form. The values $X_m$ and $j^m$ are equal to $X_0(x,t) = \nabla \beta(x,t)$; $X_1(x,t) = -\beta(x,t)\nabla v(x,t)$; $X_{i+1}(x,t) = -\nabla \beta(x,t)\mu_i(x,t)$ $(i \geq 1)$; $j^{(x)} = j_Q(x) = j`_H(x) - (u+p)p`(x)/<\rho>$; $j^1(x) = T`(x) - (\partial p/\partial u)_n H`(x)U - \sum_i (\partial p/\partial n_i)_u n_i U$; $j^{i+1}(x) = j^i_d(x) = j`_i(x) - <n_i>p`(x)/<\rho>$ $(i \geq 1)$.

(designation explained in [20]). If from $\rho^{(1)}$ (in (30)) we change to $\rho^{(2)}$ (in the formula (28)), the form of (30) will not change, but the values $F_m(x,t)$ and $X_m(x,t-y)$ will be replaced by $<F_m(x,t)>_1 = \int_0^\infty k^{(1)} \exp\{-k^{(1)}y\} F_m(x,t-y) dy$, $<X_m(x,t-y)>_1 = \int_0^\infty k^{(1)} \exp\{-k^{(1)}y_1\} X_m(x,t-y-y_1) dy_1$. As $P_m$ and $j^m$ do not depend on $t$, and for $\rho^{(r)}$ the kind of expression (30) is kept, but instead of $F_m(x,t)$ and $X_m(x,t-y)$ the values $<...<F_m(x,t)>_1...>_{r-1}$ and $<...<X_m(x,t-y)>_1...>_{r-1}$ $(r \geq 2)$ are written. It is connected with the fact that the values of $\ln\rho^{(r)}$ can be written down either from $\ln\rho^{(r-1)}$, as in (28), or likewise (30). So, expression for $\ln\rho^{(r)}$ at $\ln\rho_q$ in the form (9) have the form

$$\ln\rho^{(r)} = -<...<\Phi(t)>_1...>_{r-1} - \sum_{m=1}^n \int dx <...<F_m(x,t)>_1...>_{r-1} P_m(x) +$$

$$\int_0^\infty \exp\{-\varepsilon y\}\{d[-<...<\Phi(t-y)>_1...>_{r-1} - \sum_{m=1}^n \int dx <...<F_m(x,t-y)>_1...>_{r-1} P_m(x,-y)]/dy\} dy.$$

The expression for nonequilibrium entropy is generalized, accepting a form

$$S^{(r)} = <...<\Phi(t)>_1...>_{r-1} - \sum_{m=1}^n \int dx <...<F_m(x,t)>_1...>_{r-1} <P_m(x)>^t = -<<...<\ln\rho_q(t,0)>_1...>_{r-1}>.$$

Thus $\partial<...<\Phi(t)>_1...>_{r-1}/\partial t = -\sum_{m=1}^n \int dx [\partial(<...<F_m(x,t)>_1...>_{r-1})/\partial t] <P_m(x)>^t$ ;

$\delta<...<\Phi(t)>_1...>_{r-1}/\delta<...<F_m(x,t)>_1...>_{r-1} = -<P_m(x)>^t$;   $S^{(r)}(t) = \int S^{(r)}(x,t) dx$   ;

$\delta S^{(r)}/\delta <P_m(x)>^t = <...<F_m(x,t)>_1...>_{r-1}$;   $\Phi = \int \beta(x,t) p(x,t) dx$   ($p(x,t)$ is pressure)   ;

$$S^{(r)}(x,t) = \sum_{m=1}^n <...<F_m(x,t)>_1...>_{r-1} <P_m(x)>^t + <...<\beta(x,t)p(x,t)>_1...>_{r-1}.$$

Averaging of value $<P_m(x)>^t = \int dz \rho_q(z;t,0) \hat{P}_j(z/x) dx$ will be carried out on $\rho_q$ and coincides with averaging on $\rho^{(r)}$, $r = 1, 2, \ldots$ The entropy production operator is equal to

$\dot{S}^{(r)} = <...<\dot{S}(t-y,y)>_1...>_{r-1} = d<...<\ln\rho_q(z;t-y,y)>_1...>_{r-1}/dy =$

$d[-<...<\Phi(t-y)>_1...>_{r-1} - \sum_{m=1}^{n} \int dx <...<F_m(x,t-y)>_1...>_{r-1} P_m(x,-y)]/dy$.

The equation of entropy balance has a usual kind $\partial S^{(r)}(x,t)/\partial t = -div j^{(r)}_S(x,t) + \sigma^{(r)}(x,t)$,

where $j^{(r)}_S(x,t) = \sum_{m=1}^{n} <...<F_m(x,t)>_1...>_{r-1} <j_m(x)>^t + v(x,t)<...<\beta(x,t)p(x,t)>_1...>_{r-1}$;

$\sigma^{(r)}(x,t) = \sum_{m=1}^{n} (<j_m(x)>^t - <j_m(x)>^t_l) \nabla <...<F_m(x,t)>_1...>_{r-1} = \sum_{m=1}^{n} (<j^m(x)>^t -$

$<j^m(x)>^t_l)<...<X_m>_1...>_{r-1}$; $\partial P_m(x,t)/\partial t + \nabla j_m(x,t) = 0$; $j_0(x) = j_H(x)$; $j_1(x) = T(x)$; $J_{i+1}(x) = j_i(x)$
($T(x)$ is tensity tensor). Likewise to [20], the expressions for flows and kinetic factors are written. Other choice of thermodynamic forces is possible too. For example, as thermodynamic forces it is possible to choose Fourier-components of parameters $F_m(x,t)$ on spatial variable [20].

Let's assume that $f=2$, that is $<\Gamma^{(1)}> = <\Gamma^{(2)}>$. To justify this assumption one can note that the contribution of change of value $<\Gamma>$ is given by non-stationary processes, non-stationary periods in a history of system, when $d\bar{S}(t)/dt \neq 0$. But the averaging on $p^{(2)}(y)$ weakly influences upon properties of distribution, in which with the help $p^{(1)}(y)$ the averaging is already carried out in view of stationarity and nonstationariy intervals, allocated by averaging on $p^{(0)}(y)$. Then NESO for systems of the finite sizes differs from Zubarev's NESO [20, 21] by replacement $F_m(x,t)$ on $<F_m(x,t)>_1$, $\Phi(t)$ on $<\Phi(t)>_1$. If the first averaging (on $p^{(0)}(y)$) brings in "instantaneous" quasi-equilibrium distribution "history" of system and effects of dissipativity, one more averaging (on $p^{(1)}(y)$) brings these effects in thermodynamic parameters $F_m$, connected to thermodynamic variables $P_m$. If $r=f=2$, NESO $\rho^{(f)} = \rho^{(2)}$; $-<ln\rho^{(2)}> = S^{(2)} + \int_0^{\infty} exp\{-\varepsilon y\} << \bar{\dot{S}}(t-y,y)>_1 > dy$;

$k^{(1)} = k^{(2)}$, $<\Gamma^{(1)}> = <\Gamma^{(2)}> = 1/D^{(1)} = \{\int_0^{\infty} exp\{-k^{(1)}y\}[d<d\bar{S}(t-y)/d(t-y)>_0/dy]dy\}^{-1} = 1/k^{(1)} =$

$[\varepsilon exp\{\Delta \bar{S}^{(0)}_\varepsilon(t)\} + D^{(0)}(1-exp\{\Delta \bar{S}^{(0)}_\varepsilon(t)\})]^{-1} = [(\varepsilon - D^{(0)})exp\{(D^{(0)} + d\bar{S}^{(0)}/dt)/\varepsilon\} + D^{(0)}]^{-1}$. Then

$<<d\bar{S}(t)/dt>_0>_1 = D^{(0)} + d\bar{S}^{(0)}/dt + \varepsilon exp\{\Delta \bar{S}^{(0)}_\varepsilon(t)\} + D^{(0)}(1-exp\{\Delta \bar{S}^{(0)}_\varepsilon(t)\})$.

The entropy $\bar{S} = S^{(2)} = -<<ln\rho_q(t,0)>_1> = <\Phi(t)>_1 + \sum_{m=1}^{n} \int dx <F_m(x,t)>_1 <P_m(x)>$.

At $r=f=2$ values $<\Phi(t)>_1$, $<F_m(t)>_1 = \int_0^\infty k^{(1)} \exp\{-k^{(1)}y\} F_m(x,t-y) dy = F_m(x,t) + \int_0^\infty \exp\{-k^{(1)}y\} [dF_m(x,t-y)/dy] dy$ correspond to the "historical" description of system in view of effects having a place during its evolution, i.e. all evolution of system is considered, and, contrary to the description of works [20, 21], with the account of dissipative effects contained in $k^{(1)}$. It is possible to compare expressions for $<F_m(t)>_1$, for example, for reverse temperature, to the similar expressions obtained in extended nonequilibrium thermodynamics [53]. Let's notice that it is possible to find analogues of the relations (and relations such as (7), (28) in a general view) in the common theory of statistical estimations and statistical decisions (for example, [54]).

## 5. Concluding Remarks

The basic result of the present work consists in new interpretation of NESOM. NESO is interpreted as the average on distribution of lifetime from quasi-equilibrium distribution. It allows us to get expressions for distribution of lifetime of nonequilibrium system and moments of lifetime of system of the finite size (for infinite system the lifetime is infinite). The account for nonequilibrity as averagings on lifetime enables to write down a form of NESO for systems of the finite sizes.

Any systems are formed from elements, particles, elementary objects. The choice of these elements often represents a difficult task, in many respects determining success of the solution of a problem as a whole. It can be either the particles and quasiparticles in statistical and molecular physics, persons in a population, galaxies in cosmoloqy etc. Mathematical description of such systems is connected to concepts of stationary condition, of algorithms of phase integration of complex systems etc. In any system there are elements which enter and exit it. In case of the idealized isolated equilibrium systems the intervals between inputs tend to infinity, but the specified law remains valid. It is possible to formulate mathematically strictly the concept of lifetime of complex systems, understanding under it the time, during which there are elements in the system. It is possible to specify mathematical stochastic models, in which the lifetime is infinite. Interesting question is the problem of adequate physical interpretation of these models.

Probably, some their features peculiar to real physical systems, are responsible for such phenomena as superconductivity or superfluidity.

If one consider physical systems according to the degree of complication of the description, gas is the most simple system. The largest lifetime (at the account of all elements: both leaving, and acting), $\Gamma_0 \to \infty$, $\varepsilon \to 0$. The description of such system tends to "classical", without the account of system lifetime finiteness. A liquid is more complex system, although it, as well as gas, can flow and the exiting elements are replaced by entering ones. But in the evaporation phenomena, for example, the exit of elements is not always compensated (by condensation). The description of a liquid requires the account of finite lifetimes. The description of solid states is even more complex. In them the lifetime finiteness is essential, since the leaving elements are not replaced by new, and the system is not restored.

Summing up, the physical value of lifetime of system is introduced in consideration, as an interval of time, during which the system contains a non-zero number of elements, of which the statistical system is made. This value depends on both internal properties of system, and external influences and is generally of random character. The lifetime is connected to characteristic time intervals of system (time of collisions, time of mixing, relaxation time etc.) and with fluxes (also by temporary characteristics). So, in works [55] lifetime of finite modes of laser systems was related to the correlation time of amplitudes of modes. The average lifetime of laser modes is defined experimentally on spectra of absorption. The theory with lifetime generalizes the approaches by D.N.Zubarev [20, 21], R.L.Stratonovich [52] and extended nonequilibrium thermodynamics [53]. The analogy to internal time [11] will be carried out. The introduction of lifetime opens opportunities for strict study of a nature of time and allows to give the description of transfer processes and other nonequilibrium phenomena. The connection of value $\Gamma$, as subordinated process, with the basic random process $E$ is formally essential to the developed method. But the concept of lifetime has a deep physical sense as well, uniting Newton's approach to absolute time and idea about a matter inducing time. Parameter of lifetime contains in itself features inherent to usual dynamic variable like energy, number of particles, and coordinate variable such as time. The mathematical party of introduction of lifetime consists in obtaining of the additional information about stochastic process, except for knowledge of its stationary distribution, on stationary properties marked out by the subordinated process. The irreversibility in this approach occurs as a consequence of the assumption about existence and finiteness of lifetime of system, choosing of the moments of birth and destruction of system.

The physical values, for example, heat capacity of system, can essentially depend on lifetime. So if to assume that in system the lifetimes belonging only to some area $\Gamma_m$ can exist (it is valid, for example, for metastable states by definition), that is the contribution in measurable quantity is given by values $\Gamma \in \Gamma_m$ those areas of phase space are accessible only, where $\Gamma(E) >> t_{obs}$ (or, on the contrary, $<< t_{obs}$; instead of time of observation $t_{obs}$ there can be other characteristic scale), the average value, for example by energy, is given to expression for average energy $<E>$, dependent from $\Gamma \in \Gamma_m$. Thus as $\Gamma_m$, and expression for $<E>$ can depend on temperature either regularly, or with features, characteristic for phase transitions, when some areas of phase space suddenly become accessible (or inaccessible) for system. The dependence of $\Gamma_m$ on temperature is connected to nonequilibrium phase transitions. Detailed mesoscopic description can be derived with use of obvious stochastic models of system, for example, diffusion such as [52] or stochastic processes of a storage [14, 39].

The entropy $S$ changes at the expense of a flux acting in system entropy. In a stationary case this composed equally to entropy production in the system, the receipt of negative entropy from the outside compensates the entropy production in the system, supporting the system in a stationary state. If $\sigma_S = dS_i/dt$ is entropy production, entropy flux, arising at relaxation of system to some equilibrium state, where the system will come itself, if not perturbed, $dS_e/dt$ is a entropy flux from the outside, necessary to compensate $\sigma_S$; $dS/dt = dS_i/dt + dS_e/dt$. The value $dS_e/dt$ represents a entropy flux, caused by an exchange of both matter and energy between system and environment. At the same time for nonequilibrium processes this value can be compared with increase of the information amount at evolution of system in a direction of the most probable distribution of energy on separate components, useful effect of ordering in some part of system, flow of negentropy, determining a degree of a deviation of system from a state with maximal entropy, necessary for maintenance of a steady stationary nonequilibrium state. Thus is the entropy additivity of components of system is broken in such a manner that entropy of system becomes less than the sum entropies of its separate components. Macroscopic variables appear as the result of transition from one state in another, are determined, on the one hand, by the entropy increase (by result of energy dissipation), and, on the other hand, by the increase of the information on internal structure connected with negentropy effect.

The introduction of lifetime as thermodynamic parameter is justified by the empiric fact that real systems have finite lifetime, which essentially influences their properties and properties of their environment. The lifetime of system is represented by fundamental value having a dual nature, related to both the external time flow and to the properties of the system. The relationship between the lifetime and the nonequilibrium statistical operator method was investigated in [57], [58], [59].

**References**

[1] W. Feller, *An Introduction to Probability Theory and its Applications*, vol.2 (J.Wiley, New York, 1971).

[2] E. E. Kovalev, A. I. Vichrov and V.G. Semenov, Atomnaya energiya, *83*, (in Russian), 291, (1997).

[3] A. D. Venttsel and M. I. Freidlin, *Fluctuations in dynamic systems under action of small random perturbations* (in Russian) (Nauka, Moskow, 1979).

[4] P. Hanggi, P. Talkner and M. Borkovec, Rev. Modern Phys. *62*, 251 (1990).

[5] I. I. Gichman and A. V. Skorochod, *Introduction in the theory of random processes* (in Russian). (Izd. Nauka, Moskow, 1977).

[6] P. Gaspard and J. R. Dorfman, Phys. Rev. E, *52*, 3525 (1995).

[7] P.Gaspard, Physica A. *263*, 315 (1999).

[8] P. Salamon, A. Nitzen, B. Andersen and R. Berry, Phys. Rev. A, *21*, 2115 (1980).

[9] P. Salamon and R. S. Berry, Phys. Rev. Lett. *51*, 1127 (1983).

[10] R. L. Stratonovich, *The theory of the information* (in Russian) (Izd. "Sovetskoe Radio", Moscow, 1966).

[11] I. Prigogine, *From Being to Becoming* (Freeman, San Francisco, 1980).

[12] I. Prigogine and I. Stengers, *Entre le Temps et l'Eternite* (Fayard, Paris,1988).

[13] N. U. Prabhu, *Stochastic Storage Processes* (Springer, Berlin, 1980).

[14] V. V. Ryazanov, Ukrainian Physical Journal (in Russian) **38**, 615 (1993).

[15] N. Petrov, and I. Brankov, *Modern problems of thermodynamics* (in Russian) (Izd. "Mir", Moscow, 1986).

[16] H. Mori, I. Oppenheim and J. Ross, in *Studies in Statistical Mechanics I*, edited by J.de Boer and G.E.Uhlenbeck (North-Holland, Amsterdam, 1962), pp. 217-298.

[17] J.A. McLennan, Phys. Fluids.**4**, 1319 (1961).

[18] S. V. Peletminskii and A. A. Yatsenko, Soviet Phys JETP **26**, 773 (1968) zh. Eksp. Teor. Fiz. vol. **53**, 1327 (1967).

[19] A. I. Akhiezer and S. V. Peletminskii, *Methods of statistical physics*. (Pergamon, Oxford, 1981).

[20] D. N. Zubarev, *Nonequilibrium statistical thermodynamics*. (Plenum-Consultants Bureau, New York, 1974).

[21] D. N. Zubarev, in *Reviews of Sciense and Technology: Modern Problems of Mathematics.* Vol.15, pp. 131-226, (in Russian) ed. by R. B. Gamkrelidze, (Izd. Nauka, Moscow, 1980) [English Transl.:J. Soviet Math. **16**, 1509 (1981)].

[22] B. Robertson, Phys. Rev. **144**, 151 (1966).

[23] K. Kawasaki, J. D. Gunton, Phys. Rev. A **8**, 2048 (1972).

[24] H. Grabert, Z.Phys. B **27**, 95 (1977).

[25] R. Luzzi and A. R. Vasconcellos, Fortschr. Phys./Progr. Phys. **38**, 887 (1990).

[26] L. S. Garsia-Colin, A. R. Vasconcellos and R. Luzzi, J. Non-Equilib. Thermodyn. **19**, 24 (1994).

[27] A. R. Vasconcellos, R. Luzzi and L. S. Garsia-Colin, Phys. Rev. A **43**, 6622 (1991).

[28] A. R. Vasconcellos, R. Luzzi and L. S. Garsia-Colin, Phys. Rev. A **43**, 6633 (1991).

[29] J. G. Ramos, A. R. Vasconcellos and R. Luzzi, Fortschr. Phys./Progr. Phys. **43,** 265 (1995).

[30] J. G. Ramos and A. R. Vasconcellos, Braz. J.Phys. **27,** 585 (1997).

[31] R. Luzzi, A. R. Vasconcellos and J. G. Ramos, Braz. J.Phys. **28**, 97 (1998).

[32] R. Luzzi, A. R. Vasconcellos and J. G. Ramos, Fortschr. Phys./Progr. Phys. **47**, 401 (1999).

[33] J. G. Ramos, A. R. Vasconcellos and R. Luzzi, Fortschr. Phys./Progr. Phys. **47**, 937 (1999).

[34] D. R. Cox, *Renewal theory* (London: Methuen; New York: John Wiley, 1961).

[35] V. S. Korolyuk and A. F. Turbin, *Mathematical Foundations of the State Lumping of Large Systems* (Kluwer Acad.Publ., Dordrecht, Boston/London, 1993).

[36] R. L. Stratonovich, *The elected questions of the fluctuations theory in a radio engineering* (Gordon and Breach, New York, 1967).

[37] D. R. Cox and D. Oakes, *Analysis of Survival Data* (Chapman and Hall, London, New York, 1984).

[38] Yu. L. Klimontovich, *Statistical Theory of Open Systems* (Kluwer Acad. Publ., Dordrecht, 1995).

[39] V. V. Ryazanov, in *Physics in Ukraine*. Int. Conf., Kiev, 22-27 June,1993. Proceedings. Contributed Papers. V.5: Statistical Physics and phase transitions, pp.115-118, Kiev, Bogoliubov Institute for theoretical physics, 1993.

[40] R. Luzzi, A. R. Vasconcellos, J. G. Ramos, *A Nonequilibrium Statistical Ensemble Formalism, MaxEnt-NESOM: basic concepts, construction, application, open questions and criticism* (Los-Alamos. National Laboratory Electrone File: cond-matt - 9909160, IFGW Abstracta A 046-99. 1999).

[41] J. G. Kirkwood, J.Chem. Phys. **14**, 180 (1946).

[42] N. N. Bogoliubov, in *Studies in Statistical Mechanics I*, edited by J. de Boer and G. E. Uhlenbeck (North Holland, Amsterdam, 1962), pp. 4-118.

[43] I. Prigogine, Nature **246**, 67 (1975); Int. J. Quantum Chem. **9**, 443 (1975).

[44] C. Truesdell, *Rational Thermodynamics* (McGraw-Hill, New York, 1985), [2nd enlarged edition (Springer, Berlin, 1988)].

[45] J. Meixner, in *Irreversible Processes of Continuum Mechanics*, edited by H. Parkus and L. Sedov (Springer, Wien, 1968).

[46] J. T. Alvarez-Romero and L. S. Garsia-Colin, Physica A **232**, 207 (1996).

[47] N. G. van Kampen, in *Perspectives in Statistical Physics*, edited by H. Raveche (North Holland, Amsterdam, 1981).

[48] N. G. van Kampen, in *Fundamental Problems in Statistical Mechanics*, edited E. Cohen (North Holland, Amsterdam, 1962).

[49] J. L. del Rio and L. S. Garsia-Colin, Phys. Rev. E **54**, 950 (1996).

[50] J. A. McLennan, in *Advances in Chemical Physics* (Academic, New York, 1963),

Vol. 5, pp.261-317; *Introduction to Nonequilibrium Statistical Mechanics* (Prentice Hall, Englewood Cliffs, 1989).

[51] J. L. del Rio-Correa, L. S. Garsia-Colin, Phys Rev. E. **48**, 819 (1991).

[52] R. L. Stratonovich, *Nonlinear Nonequilibrium Thermodynamics* (Springer, Heidelberg, 1992).

[53] D. Jou, J. Casas-Vazquez and G. Lebon, *Extended Irreversible Thermodynamics* (Springer, Berlin, 1996).

[54] N. N. Chentsov, *Statistical decisive rules and optimum conclusions* (in Russian). (Izd. Nauka, Moskow, 1972).

[55] H. Atmanspacher, H. Scheingraber and C. R. Vidal, Phys. Rev. A **33,** 1052 (1986); **32**, 254, (1985).

[56] V. V. Ryazanov, *Maximum entropy principle and the form of source in nonequilibrium statistical operator method*, Preprint. Physics, arXiv:0910.4490v1.

[57] V. V. Ryazanov, Fortschritte der Phusik/Progress of Physics **49** 885 (2001).

[58] V. V. Ryazanov, Low Temperature Physics **33** 1049 (2007).

[59] V. V. Ryazanov, European Physical Journal B **72**, number 4, 629–639 (2009).